\documentclass[dvipdfmx]{iau} 
\usepackage{graphicx}
\newsavebox\myVerb

\title[Magnetic Activity in the Galactic Centre by Global MHD Simulation] 
      {Investigating Magnetic Activity in the Galactic Centre
        by Global MHD Simulation}

\author[Takeru K. Suzuki et al.]   
       {Takeru K. Suzuki$^{1,2}$, Yasuo Fukui$^2$, Kazufumi Torii$^{2,3}$,\\
         Mami Machida${^4}$, Ryoji Matsumoto$^{5}$
         \and Kensuke Kakiuchi$^2$}

       \affiliation{$^1$School of Arts \& Sciences, The University of Tokyo, \\
         3-8-1, Komaba, Meguro, Tokyo, 153-8902, Japan \\ email: {\tt stakeru@ea.c.u-tokyo.ac.jp} \\[\affilskip]
  $^2$Dept. of Physics, Nagoya University, Furo-cho, Chikusa, Nagoya, Aichi, 464-8602, Japan \\ 
  $^3$Nobeyama Radio Observatory, National Astronomical Observatory of Japan,
  462-2, Nobeyama, Minamimaki, Minamisaku, Nagano, 384-1305, Japan\\
  $^4$Dept. of Physics, Faculty of Sciences, Kyushu University, 
  6-10-1 Hakozaki, Higashi-ku, Fukuoka 812-8581, Japan \\
  $^5$Dept. of Physics, Graduate School of Science, Chiba University,
  1-33 Yayoi-cho, Inage-ku, Chiba 263-8522, Japan}

\pubyear{2016}
\volume{322}  
\jname{Title of your IAU Symposium}
\editors{A.C. Editor, B.D. Editor \& C.E. Editor, eds.}
\begin{document}

\maketitle

\begin{abstract}
By performing a global magnetohydrodynamical (MHD) simulation for the Milky Way with an axisymmetric gravitational potential, we propose that spatially dependent amplification of magnetic fields possibly explains the observed noncircular motion of the gas in the Galactic centre (GC) region. The radial distribution of the rotation frequency in the bulge region is not monotonic in general. The amplification of the magnetic field is enhanced in regions with stronger differential rotation, because magnetorotational instability and field-line stretching are more effective. The strength of the amplified magnetic field reaches $\gtrsim$ 0.5 mG, and radial flows of the gas are excited by the inhomogeneous transport of angular momentum through turbulent magnetic field that is amplified in a spatially dependent manner. As a result, the simulated position-velocity diagram exhibits a time-dependent asymmetric parallelogram-shape owing to the intermittency of the magnetic turbulence; the present model provides a viable alternative to the bar-potential-driven model for the parallelogram shape of the central molecular zone. In addition, Parker instability (magnetic buoyancy) creates vertical magnetic structure, which would correspond to observed molecular loops, and frequently excited vertical flows.  Furthermore, the time-averaged net gas flow is directed outward, whereas the flows are highly time dependent, which would contribute to the outflow from the bulge.
\end{abstract}

\firstsection 
\section{Introduction}
The magnetic field near the GC is much stronger than in the Galactic disc.
\cite[Crocker et al.(2011)]{cro11} gave a stringent lower bound $> 50$ $\mu$G on 400 pc scales from $\gamma$-ray observations.
Recently, \cite[Pillai et al. (2015)]{pil15} reported that the field strength
of a dark infrared cloud, G0.253+0.016, near the GC is $\sim$ a few mG from polarization observations (see also Pillai et al. 2016 in this volume).
These results are consistent with an inferred field strength $\sim$ mG from observed complex structure such as the nonthermal filaments \cite[(Yusef-Zadeh et al. 1984; Tsuboi et al. 1986; Lang et al. 1999; Nishiyama et al. 2010)]{yus84,tsu86,lan99,nis10}.
Such strong magnetic fields affect the dynamics of the gas in the GC region. One of the manifestations of the magnetic activity is molecular loops \cite[(Fukui et al. 2006; Fujishita et al. 2009; Torii et al. 2010a,b; Riquelme et al. 2010; Kudo et al. 2011)]{fuk06,fuj09,tor10a,tor10b,riq10,kud11},
which are supposed to 
rise upwards as a result of magnetic buoyancy \cite[(Parker instability; Parker 1966)]{pak66}.
Japanese groups \cite[(e.g., Shibata \& Matsumoto 1991; Nishikori et al. 2006; Machida et al.2009; 2013)]{sm91,mac09,mac13} have investigated magnetic activity in the Galaxy with MHD simulations.
Following these works, we \cite[(Suzuki et al. 2015; S15 hereafter)]{suz15} performed a global MHD simulation, particularly focusing on the magnetic activity in the Galactic bulge region.
In this proceedings paper, we introduce the results of our global simulation based on S15.

\section{Simulation Setup \& Overview}
\begin{figure}[h]
\begin{center}
  \includegraphics[width=3.4in]{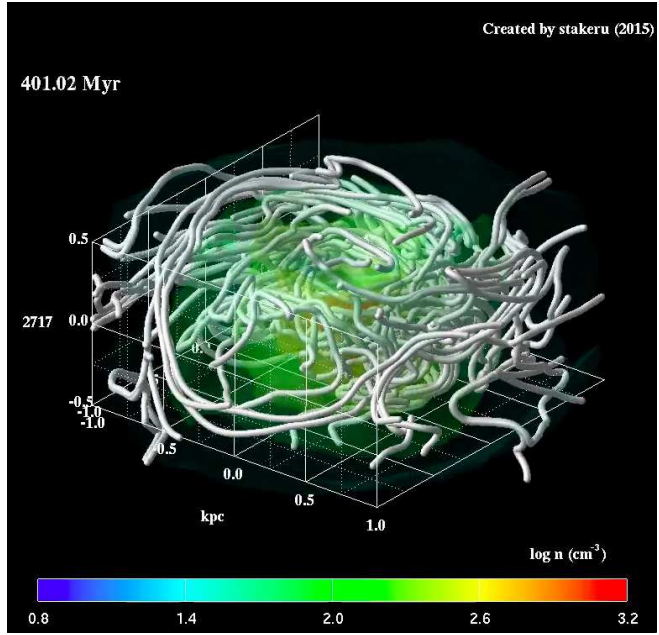}
 \caption{Snapshot view of the bulge region $<$ 1 kpc at $t=401.0$ Myrs. Colours indicate
   number density, $n$ cm$^{-3}$, in logarithmic
   scale, white lines denote magnetic field lines. }
 \label{fig:overview}
\end{center}
\end{figure}

Our MHD code was originally developed for simulating protoplanetary discs \cite[(Suzuki \& Inutsuka 2009; 2014; Suzuki et al.2010)]{si09,si14,suz10}.
In order to simulate the Galactic bulge and disc, we modified the external gravity to the {\it axisymmetric} Galactic potential model introduced by \cite[Miyamoto \& Nagai (1975)]{mm75}.
We initially set up a very weak vertical magnetic field with the plasma $\beta$ ($=$gas pressure/magnetic pressure) value of $10^5$ at the Galactic plane. 
We would like readers to refer to S15 for the detail of the simulation setup. 

We need to note that in S15 the conversion from 
dimensionless units for the simulation to 
physical units contains an error and the time unit in S15 must be $\approx 10\%$ smaller.
For instance, Fig. 3 of S15 presents snapshots at $t=439.02$ Myrs, but the time must be corrected to 
$t=401.0$ Myrs as shown in Fig. \ref{fig:overview}.
Fig. \ref{fig:overview} illustrates moderately turbulent magnetic field that is a result of the amplification of the initial weak magnetic field by the magnetorotational instability \cite[(Velikhov 1959; Chandrasekhar 1961; Balbus \& Hawley 1991)]{vel59,cha61,bh91}, Parker instability \cite[(Parker 1966)]{pak66}, and field-line stretching 
by differential rotation
\begin{lrbox}\myVerb
  \scriptsize{\verb|http://ea.c.u-tokyo.ac.jp/astro/Members/stakeru/research/galaxypot/glbdsk28inv_lagp2.mp4| }
\end{lrbox}
\footnote{Movie of the simulation is available at \\ \usebox\myVerb}.   
In Fig. \ref{fig:overview}, we select different field lines from Fig. 3 of S15, and readers may notice that the appearances are different although the same snapshot data is used.

\section{Results \& Discussion}

\begin{figure}[h]
\begin{center}
 \includegraphics[width=3.in]{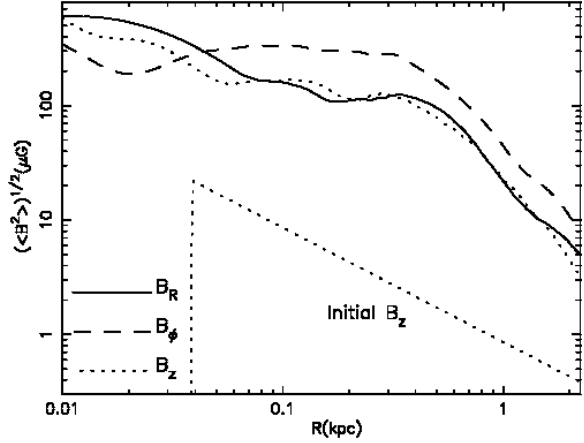} 
 \caption{Radial distribution of the time-averaged magnetic field strength. Solid, dashed, and dotted lines correspond to the radial, azimuthal, and vertical components, respectively. }
   \label{fig:B}
\end{center}
\end{figure}

Figure \ref{fig:B} presents the radial profile of each component of the time-averaged magnetic field strength after the amplification of the magnetic field is saturated. 
In the almost entire region, $>40$ pc, the toroidal component ($B_{\phi}$) dominates the poloidal component ($B_R$ and $B_z$) because the differential rotation efficiently amplifies the magnetic field by the stretching motion.
However, the poloidal component is not so weak and the total field strength,
$\sqrt{B_R^2 + B_{\phi}^2 + B_z^2}$, reaches $\sim 0.5$ mG inside $\lesssim 0.5$ kpc.

\begin{figure}[h]
 \vspace*{-0.3 cm}
\begin{center}
\hspace{-2.cm} \includegraphics[width=3.3in]{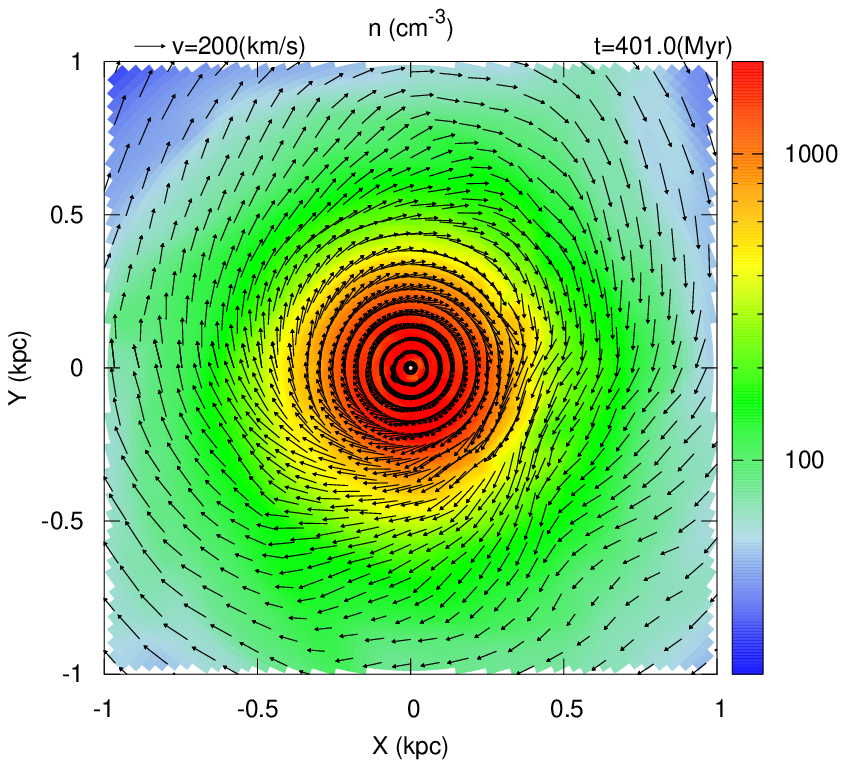} 
\hspace{-1.5cm}\includegraphics[width=3.3in]{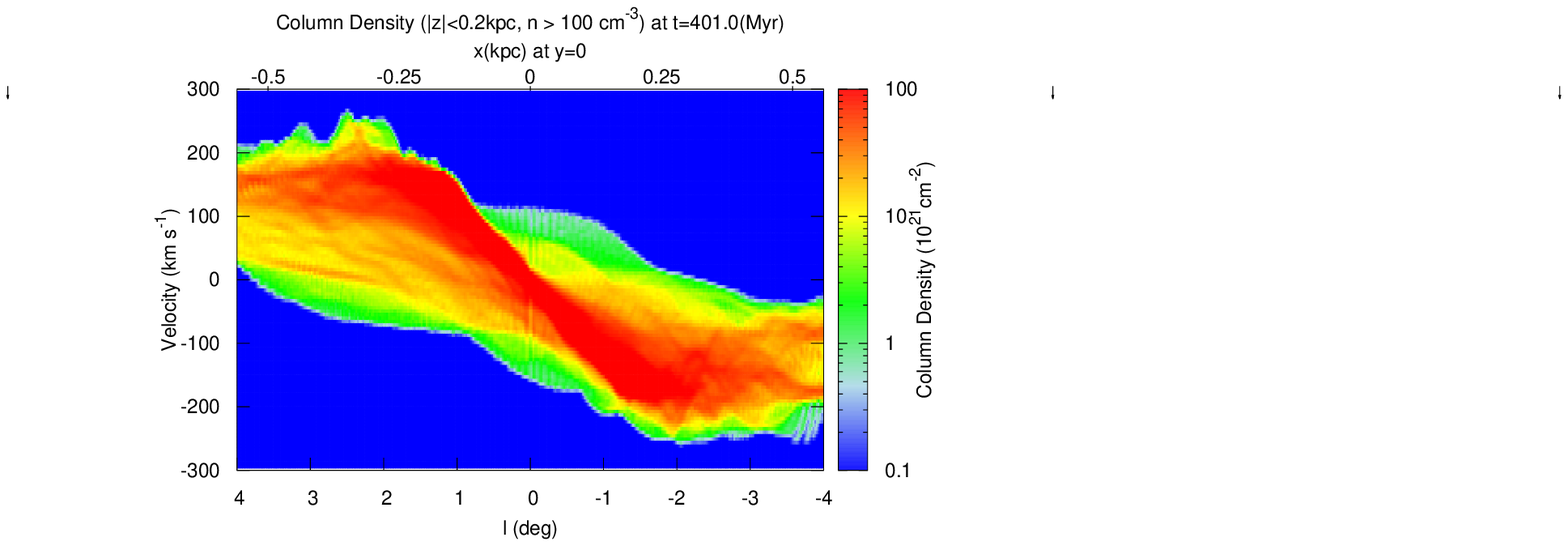}
\vspace*{-0.9 cm}
 \caption{{\it left:} Face-on views of density in units of $n$ cm$^{-3}$ (colour) and velocity field (arrows) at the Galactic plane. {\it Right:} Simulated $l-v$ diagram. Colours indicate column density integrated along line of sight.  }
   \label{fig:faceon}
\end{center}
\vspace*{-0.2cm}
\end{figure}

The left panel of Fig. \ref{fig:faceon} shows density (colours) and velocity (arrows) fields at the Galactic plane \begin{lrbox}\myVerb
  \scriptsize{\verb|http://ea.c.u-tokyo.ac.jp/astro/Members/stakeru/research/galaxypot/faceon28inv_4.gif| }
\end{lrbox}
\footnote{Movie is available at \\ \usebox\myVerb}.
The velocity field shows radial flows, in addition to the background rotating component.
These radial flows are excited by the inhomogeneous transport of the angular momentum.
The angular velocity of the rotation is not smoothly distributed with radius because the contribution from the radial pressure gradient force is spatially dependent.
MHD turbulence, which transports the angular momentum outward, is developed more effectively in regions with stronger differential rotation.
Therefore, angular momentum is transported more efficiently for stronger differential rotation.
As a result, the radial force balance of gravity, centrifugal force, and pressure gradient force breaks down, and the fluid element moves radially inward or outward. 
In addition to this mechanism, the inhomogeneous gradient of magnetic pressure also drives radial flows.

Radial (noncircular) flows are also observed as a characteristic feature in a position -- velocity diagram.
The right panel of Fig. \ref{fig:faceon} shows column density in a plane of Galactic longitude, $l$, vs. line-of-sight velocity, $v$.
One can see a ``parallelogram'' shape in the central part of the diagram, which can be directly compared to the observed ''asymmetric parallelogram'' \cite[(e.g., Liszt \& Burton 1980; Bally et al. 1987; Takeuchi et al. 2010)]{lb80,bal87,tak10}.
The shape of the simulated parallelogram changes with time
\begin{lrbox}\myVerb
  \scriptsize{\verb|http://ea.c.u-tokyo.ac.jp/astro/Members/stakeru/research/galaxypot/pv-anim28inv_hr_2.gif| }
\end{lrbox}
\footnote{Movie is available at \\ \usebox\myVerb},
depending on regions in which fast radial flows are excited. 

In our simulation vertical flows are also frequently excited by magnetic activity, and such flows contribute to the outflow from the Galactic bulge region (see Kakiuchi et al. 2016 in this volume for the detail).

This work was supported by Grants-in-Aid for Scientific Research from the MEXT of Japan, 24224005 (PI: YF).
Numerical simulations in this work were carried out at the Cray XC30 (ATERUI) operated in CfCA, National Astrophysical Observatory of Japan, and the Yukawa Institute Computer Facility, SR16000.

\nocite{*}
\bibliography{IAU32201}

\end{document}